\documentclass[
reprint,
superscriptaddress,
showpacs,
preprintnumbers,
 amsmath,amssymb,
 aps,
]{revtex4-1}

\usepackage{graphicx}
\usepackage{dcolumn}
\usepackage{bm}
\usepackage[T1]{fontenc}
\usepackage{textcomp}

\usepackage{amsfonts}
\usepackage{upgreek}     
\usepackage[T1]{fontenc} 
\usepackage{ae,aecompl}  

\newcommand{\micron}{~\ensuremath{\upmu\text{m}}}
\newcommand{\microsec}{~\ensuremath{\upmu\text{s}}}
\newcommand{\umns}{\ensuremath{\upmu\text{m}}}
\newcommand{\usns}{\ensuremath{\upmu\text{s}}}

\begin{document}

\title{Measuring and manipulating the temperature of cold molecules trapped on a chip}

\author{S. Marx}
\author{D. Adu Smith}
\affiliation{Fritz-Haber-Institut der Max-Planck-Gesellschaft, Faradayweg 4-6, 14195 Berlin, Germany}
\author{B. Sartakov}
\affiliation{A.M. Prokhorov General Physics Institute, RAS, Vavilov Street 38, Moscow 119991, Russia}
\author{G. Meijer}
\affiliation{Fritz-Haber-Institut der Max-Planck-Gesellschaft, Faradayweg 4-6, 14195 Berlin, Germany}
\affiliation{Radboud University of Nijmegen, Institute for Molecules and Materials, Heijendaalseweg 135, 6525 AJ Nijmegen, The Netherlands}
\author{G. Santambrogio}
\affiliation{Fritz-Haber-Institut der Max-Planck-Gesellschaft, Faradayweg 4-6, 14195 Berlin, Germany}
\affiliation{Istituto Nazionale di Ottica CNR and LENS, Via N. Carrara 1, 50019 Sesto Fiorentino, Italy}
\email{gabriele.santambrogio@fhi-berlin.mpg.de}

\date{\today}

\begin{abstract}

We demonstrate the measurement and manipulation of the temperature of cold CO molecules in a microchip environment.
Through the use of time-resolved spatial imaging, we are able to observe the phase-space distribution of the molecules, and hence deduce the corresponding temperature. We do this both by observing the expansion of the molecular ensemble in time and through the use of numerical trajectory simulations. Furthermore, we demonstrate the adiabatic cooling of the trapped molecular sample and discuss this process.
\end{abstract}

\maketitle

Cold and ultracold molecules are gaining ever more attention due to their potential for studying new physical and chemical phenomena, such as ultracold chemistry, fundamental symmetry tests, quantum information and quantum simulation.\cite{Carr_NewJPhys11p055049_2009}
The realization of such molecular ensembles has seen great progress via the binding of ultracold atoms, through which, for example, rovibronic ground-state molecules have been realised~\cite{Danzl_NatPhys6p265_2010} and quantum-state-specific chemical reactions have been observed and controlled~\cite{Ospelkaus_Science327p853_2012}.
On the other hand, molecular beam experiments have demonstrated significant progress in the capture and control of  cold samples of molecules that are not constituted of laser-cooled atoms, for example, O$_2$~\cite{Narevicius_PhysRevA77p051401_2008}, OH~\cite{Meerakker_PhysRevLett94p023004_2005} and ND$_3$~\cite{Bethlem_PhysRevA65p053416_2002}, as well as CH$_3$F, CF$_3$H, and CF$_3$CCH~\cite{Chervenkov_PhysRevLett112p013001_2014}.
Experiments have also now demonstrated the direct laser-cooling of molecules~\cite{Shuman_Nature467p820_2010,Hummon_PhysRevLett110p143001_2013} and a three-dimensional magneto-optical trap~\cite{Barry_arXiv1404.5680_2014}. Furthermore, forced evaporative cooling of cold OH molecules has also been reported~\cite{Stuhl_Nature492p396_2012}.

\begin{figure}
\centering
\includegraphics[width=.48\textwidth]{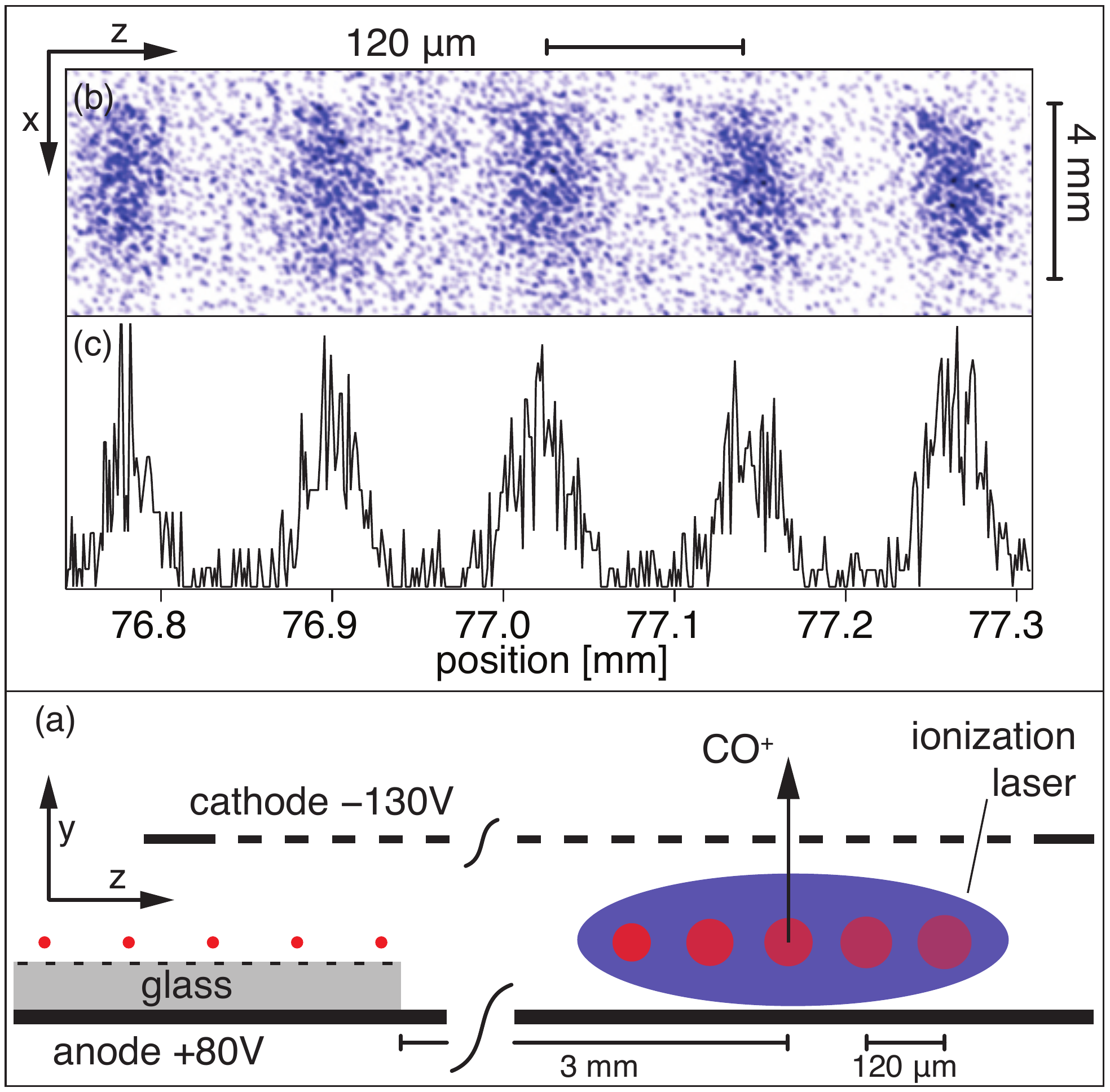}
\caption{(a) Detection region of the chip. Molecules trapped above the microelectrode array (red dots) are released from the traps at a well-defined velocity, whereupon they travel into the detection region of the chip. The molecules are ionized using the REMPI process and then propelled by the electric field created between the anode and the cathode (a ring electrode) to the ion lenses (not shown), which image the ion spatial distribution on an MCP with phosphor screen. A CCD camera records the resultant image. The microtraps are spaced by 120\micron\ and the diagram is not to scale; the actual distances are marked for reference. (b) Example 2D image of molecules released from an array of microtraps.\cite{Marx_PhysRevLett.111_243007} (c) Integrated line profile of image in (b).
\label{fig:SetupImaging}} 
\end{figure}

A promising tool for the control and manipulation of cold molecules is the molecule chip~\cite{Meek_Science324p1699_2009,Englert_PhysRevLett107p263003_2011,Marx_PhysRevLett.111_243007}, the molecular analogue of the atom chip~\cite{Hansel_Nature413p498_2001,Folman2002_263,Fortagh_RevModPhys79p235_2007,Reichel:AtomChip} or ion chip~\cite{Stick_NatPhys2p36_2006,Ospelkaus_Nature476p181_2011}.
Using the molecule chip we have recently demonstrated the integrated on-chip time-resolved spatial imaging of cold molecules, in a manner that is both quantum state selective and generally applicable.\cite{Marx_PhysRevLett.111_243007}
One straightforward application of this new capability is to image the spatial distribution of a molecular ensemble and, by taking images at different times, to access the phase-space distribution.  We show here that we can use these methods to estimate the temperature of a cold trapped molecular ensemble.
Our imaging technique is similar in its resulting images to time-of-flight imaging techniques used for ultracold atoms.\cite{Ketterle_ProcFermi_1999,Bucker_1367-2630-11-10-103039,Ockeloen_PhysRevA.82.061606,Smith_OptExpress19p8471_2011}
We observe the time-of-flight expansion of the molecular ensemble and use it to determine the temperature and also show that the deeper the molecule chip trap, the higher the observed temperature.
Moreover, we use imaging on the chip to investigate a manipulation sequence that adiabatically cools the trapped molecular ensemble.

The experimental setup we use here was first described in Ref.~\cite{Marx_PhysRevLett.111_243007}. Here we provide only the most important information relevant to the measurement of the temperature. We define a right-handed coordinate system in which the $z$ direction is oriented in the propagation direction of the molecular beam and the $y$ direction is normal to the molecule chip surface. 

Our molecule chip creates an array of tubular microtraps for polar molecules in low-field-seeking states. Each trap has a diameter of approximately 20\micron\ with its axis approximately 25\micron\ above the chip substrate, is 4~mm long in the $x$ direction and can be moved at will in the $z$ direction, i.e.\ along the molecular beam direction. 

We produce a packet of cold carbon monoxide ($^{13}$CO) molecules in the upper component of the $a^3\Pi_1$, $v=0$, $J=1$ $\Lambda$-doublet by intersecting a supersonic molecular beam with a 10~ns laser pulse at 206~nm (150~MHz bandwidth, 0.5 mJ)~\cite{Velarde_RevSciInstr81p063106_2010}. The molecules are then subsequently loaded directly on the chip from the molecular beam by capturing them in the microtraps that are initially made to move at the same speed as the molecular beam. Typically we fill a series of about 10 microtraps. 
Immediately after loading the molecules on the chip, the microtraps are decelerated to 138~m/s applying an acceleration of $10^6$~m/s$^2$ (1\micron/\usns$^2$). 
This deceleration is sufficient to separate the trapped molecules from the background of untrapped molecules.

For imaging detection (Fig.~\ref{fig:SetupImaging}), the molecules are released from the microtraps in the $z$ direction so that they can expand ballistically for a tunable time duration to allow for interrogation of their phase-space distribution in the $x$-$z$ plane~\cite{Marx_PhysRevLett.111_243007}.
The release of the microtraps occurs sequentially: upon arrival at the end of the microtrap array each trap rapidly opens out within hundreds of nanoseconds (i.e. instantaneously for the molecules).
Using a (1+1) REMPI process~\cite{Ashfold_AnnuRevPhysChem45p57_1994}, the molecules are ionized via the $b^3\Sigma^+$, $v=0$, $N=1$ state using 0.8~mJ/mm$^2$ of laser light at 283~nm~\footnote{The 283~nm light is obtained by frequency doubling the output of a Radiant Dyes NarrowScan laser pumped by the second harmonic of a Nd:YAG laser.} that propagates parallel to the chip surface. 
The ionization takes place between two parallel electrodes, an anode and a cathode, that guarantee the field homogeneity necessary for imaging~\cite{Marx_PhysRevLett.111_243007}. 
The anode is recessed 2~mm below the plane of the microtraps to allow space for the ballistic expansion of the molecular ensemble (Fig.~\ref{fig:SetupImaging}).
A standard set of ion lenses is then used to image the CO cations onto a MCP detector with phosphor screen situated 40~cm above the chip surface~\cite{Marx_PhysRevLett.111_243007}. 
A CCD camera is used to record the image.

\begin{figure}
\centering
\includegraphics[width=.48\textwidth]{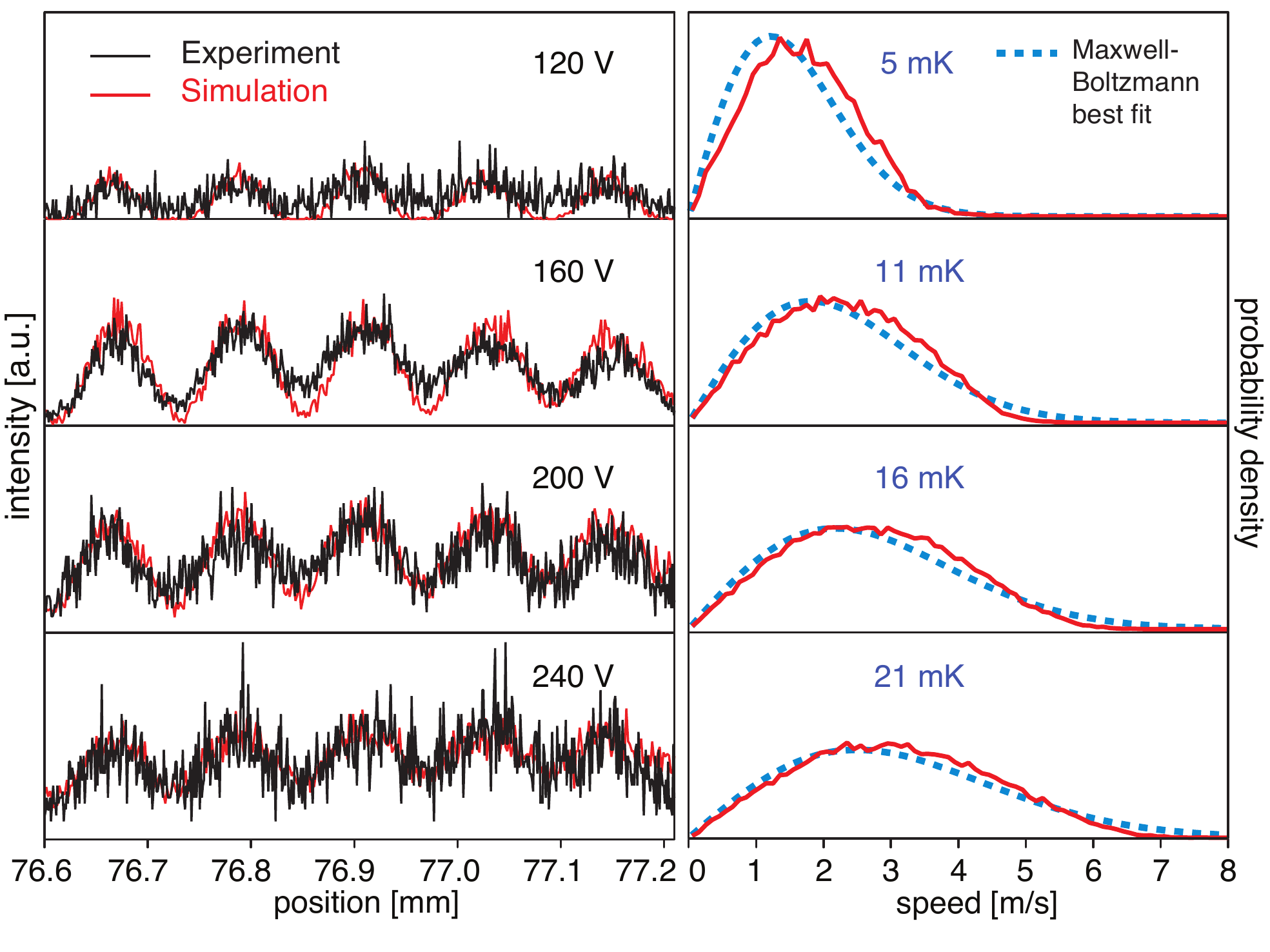}
\caption{\textit{left:} Integrated line profiles (black) from images of molecules for differing microelectrode voltages (i.e.~differing trap depths), along with the results of numerical trajectory simulations (red). The maximal relative speed $v_\text{max}=\sqrt{2\,U/m}$ of stably trapped molecules (given by the trap depth $U$) was 2.4, 4.0, 5.1, and 6.1~m/s, respectively. \textit{right:} Corresponding speed distributions (red) extracted from the trajectory simulations, along with the best-fit Maxwell-Boltzmann curve (blue dashed), labelled with the best-fit temperature. Speeds are given relative to the mean forward velocity of the molecular cloud.
\label{fig2:voltages}} 
\end{figure}

An example image of molecules is shown in Fig.~\ref{fig:SetupImaging}(b).
This is the sum of approximately 10$^5$ experimental cycles. 
The dynamics of the molecules along the 4-mm length of the microtraps ($x$ direction) is negligible for the experiments presented here because the molecules almost never experience a force in that direction during the relatively short time they spend on the chip. 
We therefore integrate the signal along the $x$ direction (vertical axis of the images) and concentrate on the perpendicular direction, as shown in  Fig.~\ref{fig:SetupImaging}(c).
For each individual molecular cloud, the distance between release from the microtrap and detection is fixed. 
We therefore control the expansion time by controlling the velocity at which the molecular clouds are ejected from the microtraps, i.e. by defining the speed at which the microtraps move over the chip surface.  
The ballistic expansion times given later in the paper are thus for the central cloud in each image. 
Within the signal-to-noise of our data, any difference in cloud size between the rightmost and leftmost clouds (due to slightly differing expansion times) was undetectable (see, for example, Fig.~\ref{fig:SetupImaging}(b, c)). 

The excitation laser (206~nm) has a spot size of roughly 1~mm and by the time the metastable molecules reach the chip's entrance 40~mm downstream, their phase-space distribution shows a strong correlation between position and velocity in the $z$ direction. 
The faster molecules have been moving toward the front of the packet while the slower have been lagging behind, so that by the time the packet reaches the chip's entrance it is roughly 4~mm long and its local average velocity in the $z$ direction changes by 9~m/s every mm. 
However, over the 20-\umns\ size of each microtrap, any correlation between position and velocity is negligible and we can assume a uniform distribution in phase-space: The  distribution of the captured molecules is limited in all directions by the acceptance of the microtraps, except for the velocity component in the $y$ direction, for which the microtraps are slightly under-filled. 

The bottom of each microtrap can be approximated by a harmonic potential, the flanks are rather flat and there is a saddle point in the $y$ direction when the microtrap is in uniform motion~\cite{Meek_NewJPhys11p055024_2009}.
A fraction of molecules are trapped in metastable trajectories and are unable to find a way out through the saddle point during the duration of our experiments: however, from previous studies~\cite{Meek_NewJPhys11p055024_2009} we know that the number of molecules captured with metastable trajectories are a small percentage of the total number of molecules.
Under uniform acceleration, the microtrap becomes shallower and its shape is rotated in the $y$-$z$ plane.\cite{Meek_NewJPhys11p055024_2009}

The molecules spend more than 200\microsec\ in the microtraps, which have typical frequencies in the range of hundreds of kHz for the $y$ and $z$ directions, before being released and imaged.
Upon release, each molecule has thus oscillated tens of times in the microtrap's mechanical potential in the $y$ and $z$ directions (but experiencing practically no force in the $x$ direction).

The most insightful method of investigating the phase-space distribution inside the microtraps involves carrying out numerical trajectory simulations and comparing their results with the experimental measurements. 
We have shown that simulations are in good agreement with experiments.\cite{Marx_PhysRevLett.111_243007}
The low number density (10$^7$/cm$^3$) rules out any thermalization of the sample.
However, when observing the velocity distributions given by the trajectory simulations, we find that they approximate very closely Maxwell-Boltzmann distributions~\cite{Marx_PhysRevLett.111_243007}.
We stress that this is due to the shape of the microtraps and not to any sort of thermalization process.
Hence, although the temperature is not strictly defined, the characterization of our molecular ensemble using a temperature is useful and describes what we observe very closely.

For example, Fig.~\ref{fig2:voltages}(a) shows four different measurements of molecular distributions after trapping with microtraps of different depth but otherwise similar shape. 
The depth of the microtraps is controlled by the amplitude of the voltage waveforms applied to the microelectrodes on the chip surface, which for these measurements was, respectively, 120, 160, 200, and 240~V. 
After the initial deceleration phase to separate the trapped molecules from the background gas, the microtraps were made to move uniformly for the final phase of the manipulation sequence.
As the traps are shallower when decelerated, the depth under deceleration is what defines the phase-space acceptance of the experiment. 
Both from an analytical description of the electric field of the microtraps and from finite element simulations, we know the trap depth for any given amplitude of the applied voltage waveforms: 10, 28, 46, and 65~mK, respectively, under deceleration and 39, 55, 71, and 87~mK, respectively, under uniform motion.
The trajectory simulations reproduce the experiments well and, as shown in Fig.~\ref{fig2:voltages}(b), the resultant speed distributions are best fit with temperatures of 5, 11, 16, and 21~mK, respectively.

\begin{figure}
\centering
\includegraphics[width=.48\textwidth]{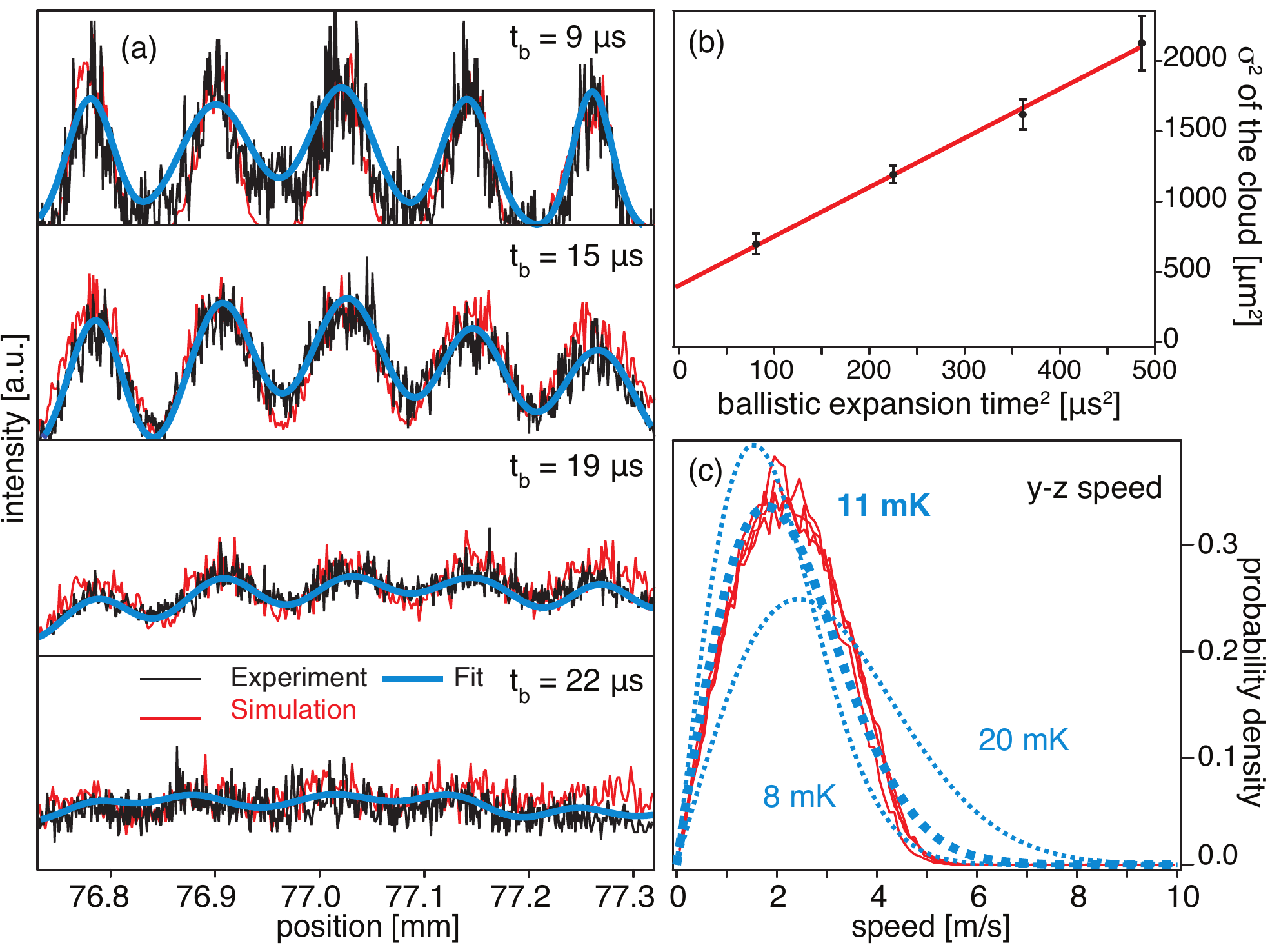}
\caption{(a) Integrated line profiles (black) extracted from images (as in Fig.~\ref{fig:SetupImaging}) for various expansion times. For each expansion time, the blue line is the result of fitting a multi-gaussian profile (see text for details), the red line is the result of trajectory simulations. (b) The square of the mean gaussian standard deviation from the fit in (a) plotted against the ballistic expansion time squared. The slope is proportional to the temperature of the gas (Eq.~(\ref{eq:expansion})). (c) Speed distributions calculated from trajectory simulations for all four experimental conditions (red) and calculated Maxwell-Boltzmann distributions (blue), with the best fit in bold.
\label{fig:analytic_temp}} 
\end{figure}

The above results show that we can trap a molecular ensemble with a given temperature, defined by choosing the depth of the microtraps on the molecule chip, and that we can model the system accurately using numerical trajectory simulations.  However, here we have not yet used the ability to take snapshots at different times in the ballistic time-of-flight evolution of the molecular ensemble.  We therefore now repeat the experiment at an electrode voltage of 160~V but this time we record multiple images during the ballistic expansion of the molecule cloud. In Fig.~\ref{fig:analytic_temp}(b) the integrated line profiles are shown after ballistic times-of-flight of 9, 15, 19, and 22\microsec. The ballistic expansion of each individual molecular cloud (each from an individual microtrap) can be seen with increasing expansion time. However, for times $>20$\microsec\, it becomes increasingly difficult to discern the individual microtraps as the individual clouds expand into one another.  It is for this reason that integrated on-chip imaging is important: longer times of flight to an external detector would see the spatial structure being completely washed out. For the expansion times of 9, 15, 19, and 22\microsec\ (over a fixed distance of 3~mm) the molecules were released when traveling uniformly at 336, 207, 162, and 138~m/s, respectively. We took care that the molecules experienced the same trap depth and shape for each measurement, since observing the evolution of the system is only useful if the initial conditions are the same for each measurement.\cite{Marx_PhysRevLett.111_243007}

This method of time-of-flight imaging (i.e.~where the expansion of a gas is monitored over time after release from a trap) has been very successful in determining the temperature of cold atomic gases~\cite{Weiss:89_JOptSocAmB}. In the atomic case, an atom cloud is illuminated with a laser beam tuned to a closed optical transition. To gain an image of the cloud, either the many scattered photons are imaged (fluorescence imaging) or the shadow cast in the laser beam is imaged (absorption imaging).\cite{Ketterle_ProcFermi_1999,Bucker_1367-2630-11-10-103039,Ockeloen_PhysRevA.82.061606,Smith_OptExpress19p8471_2011} The expansion of the gas over time is related to the temperature of the gas and hence this method is a relatively straight-forward way of ascertaining the temperature.
We want to apply here the same analysis, but, of course, for cold molecules.  If the expansion of the atomic or molecular ensemble is dominated by the translational temperature, i.e. the velocity distribution of the particles, then the expansion can be described as~\cite{Weiss:89_JOptSocAmB}:
\begin{equation}
\sigma^2(t_b) = \sigma_{i}^2 + \frac{k_{B}T}{m}t_b^2,
\label{eq:expansion}
\end{equation}
where $\sigma$ is the cloud standard deviation at ballistic expansion time $t_b$, $\sigma_{i}$ is the initial cloud standard deviation, $m$ is the mass of the particle and $T$ is the temperature.  This analysis functions on the premise that the clouds are gaussian in form both in their position and velocity distributions~\cite{Weiss:89_JOptSocAmB}. However, if $\sigma_{i} \ll \sigma(t_b)$, Eq.~\eqref{eq:expansion} remains a good approximation even in the case when the initial spatial distribution is not gaussian.

To apply this analysis to our time-resolved images of molecules expanding from the molecule chip trap, we fit each of the line profiles in Fig.~\ref{fig:analytic_temp}(a) with a sum of seven gaussian functions (also plotted in Fig.~\ref{fig:analytic_temp}(a)), including the five clouds seen in the image and the contributions from their next-nearest neighbors on either side of the image. Using Eq.~\eqref{eq:expansion}, we then carry out a least-squares fit to $\sigma^{2}(t_b)$ against $t_b^2$ and we subsequently extract a temperature of $T = 12 \pm 2$~mK, where the error is given here by the 95\% confidence bounds of the least-squares fit (see Fig.~\ref{fig:analytic_temp}(b)).  This compares well with the 11 mK found using trajectory simulations (Fig.~\ref{fig2:voltages}).

\begin{figure}
\centering
\includegraphics[width=.49\textwidth]{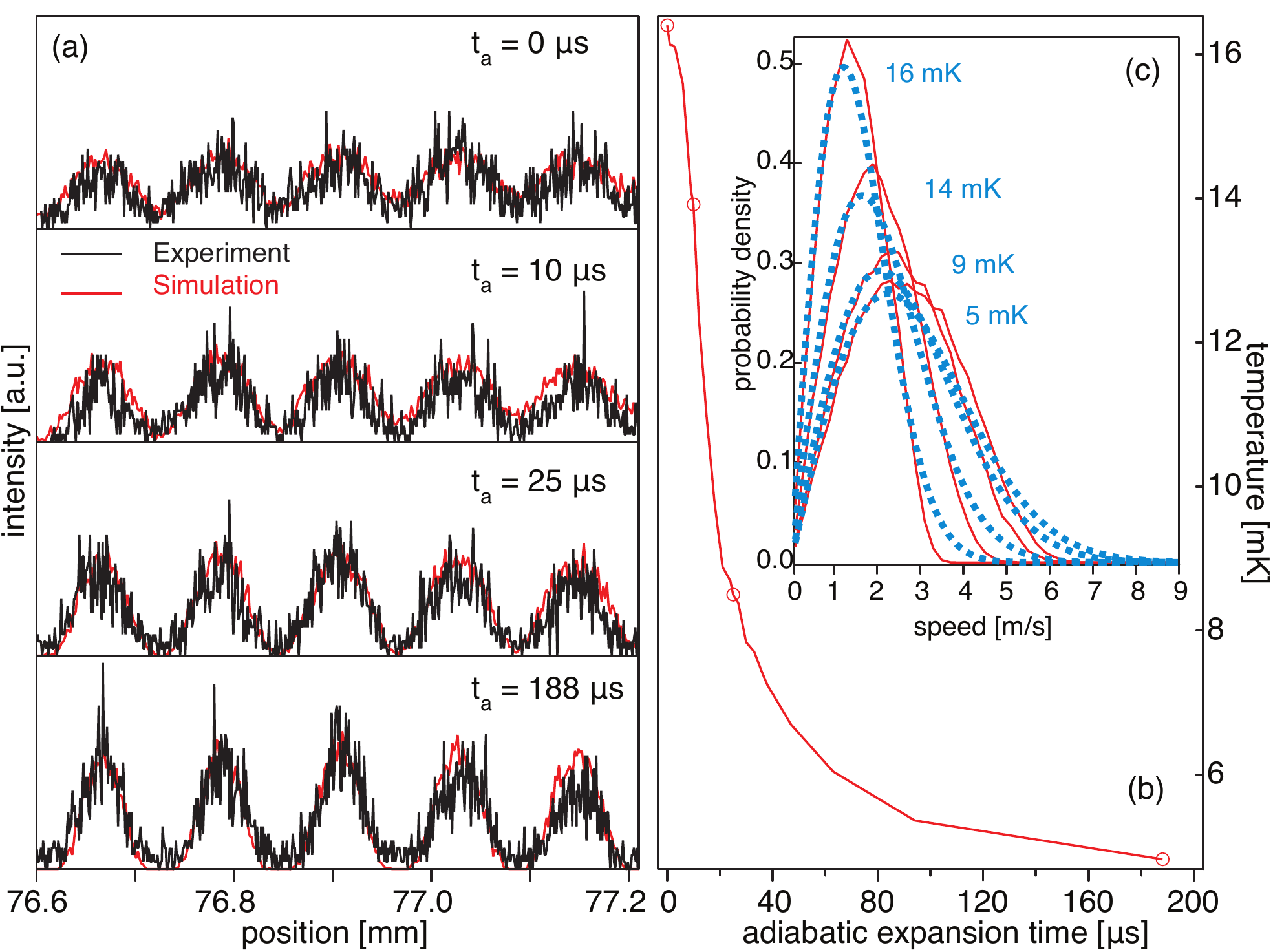}
\caption{(a) Experimental integrated line profiles of molecules for various manipulation times and along with corresponding numerical trajectory simulations. (b) Decrease in temperature with manipulation time from trajectory simulations. Circles denote the times for which experimental measurements were performed. (c) Change in the speed distribution with manipulation time. All temperatures refer to the best Maxwell-Boltzmann fits (blue) to the calculated distributions (red), with the standard deviation of the temperature fit parameter $T$ being in each case approximately 3\% of $T$. 
\label{fig:cooling}} 
\end{figure}

We have recently shown that we can adiabatically cool the trapped molecules with an expansion of the trapping potential~\cite{Marx_PhysRevLett.111_243007}.
To do this experimentally, we capture and decelerate molecules using waveforms with 200~V amplitude. 
We then ramp down the amplitude of the waveforms linearly to 50~V in a time $t_a$ while guiding the molecules at constant velocity over the molecule chip surface. This procedure expands the volume of the traps in the $y$ and $z$ directions and the trap depth is lowered from 71~mK to 13~mK.
Figure~\ref{fig:cooling} shows integrated experimental imaging signals along with corresponding trajectory simulations for $t_a = 0$, 10, 25, and 188\microsec. 
The data show that the best-fit temperature is reduced from 16~mK to 5~mK for an expansion time $t_a=188$\microsec.
For shorter expansion times, however, cooling is less effective, as can be seen from Fig.~\ref{fig:cooling}(b). 

A simple way to rationalize the results of experiments and trajectory simulations is to approximate the trapping potential with a harmonic one, so that it becomes $U=k(y^2+z^2)/2$.
The lowering of the trapping potential for the adiabatic cooling corresponds therefore to a reduction of the initial $k_i$ to a final $k_f$, which results in a reduction of the trap frequency $\omega=\sqrt{k/m}$, where $m$ is the mass of a molecule. 
If the transformation is adiabatic, the total energy of the oscillator remains proportional to the frequency~\cite{Landau-Lifshitz:Mechanics1993}. 
Therefore, the final energy of our ensemble will be given by  $E_f=E_i\, \omega_f/\omega_i=E_i \sqrt{k_f/k_i}$.
Furthermore, if the system is treated quantum mechanically, the energy is given by $E = (n+\frac{1}{2}) \hbar \omega$ and the adiabaticity condition implies that each molecule remains in the same $n$-level during the process.
This gives the same dependence of the energy change on the trapping potential as in the classical case. 

As we mentioned above, the microtrap potentials on the chip are not harmonic over the whole spatial extent of the microtrap. 
We therefore take the central 10\micron\ of the potential (where the vast majority of molecules are situated) and fit a harmonic function $U=(k_{y}y^2+k_{z}z^2)/2$. 
This gives a trapping frequency of approximately $1300$~kHz at 200~V and approximately $500$~kHz at 50~V, which leads to a reduction in temperature to around 40\% of the initial temperature, i.e. from 16~mK to 6~mK. 
The main source of error in this treatment is the harmonic approximation of the trapping potential. 
Moreover, the rate of change of the trapping potential, and therefore of the trap frequency, must be slow enough for the process to be adiabatic: $d\omega / dt \ll \omega^2$.\cite{Landau-Lifshitz:Mechanics1993}  As the oscillation period is $\mathcal{T}=2\pi/\omega$, one can rewrite the adiabaticity condition as $d\mathcal{T}/dt \ll 1$.  In our case, the initial trap period was 0.8\microsec\ (for 1300~kHz) and the final trap period was 2.0\microsec\ (for 500~kHz).  Taken simply as a change in trap period of $\Delta \mathcal{T} = 1.2$\microsec\ in the adiabatic expansion time of 10, 25, and 188\microsec\ (Fig.~\ref{fig:cooling}), leads to $\Delta \mathcal{T} / \Delta t_a =$ 0.1, 0.05, and 0.007, respectively. One can see the validity of the adiabatic condition in Fig.~\ref{fig:cooling}(b), where only in the latter case ($t_a = 188$\microsec), does the temperature approach the asymptotic limit, i.e. when $\Delta \mathcal{T} / \Delta t_a = 0.007 \ll 1$. 

With our ability to trap molecules~\cite{Meek_Science324p1699_2009}, manipulate their internal~\cite{Santambrogio_ChemPhysChem12p1799_2011,Abel_MolecularPhysics110p1829_2012} and external~\cite{Meek_PhysRevLett100p153003_2008,Meek_NewJPhys11p055024_2009} degrees of freedom and now produce time-resolved images with a fully integrated detection system~\cite{Marx_PhysRevLett.111_243007}, the molecule chip is being developed into a complete toolkit for the investigation of cold molecular ensembles.  We have shown here that this toolkit can be used to measure the temperature of the trapped molecules through time-of-flight imaging. Using a sequence of time-resolved images, the free expansion of the molecular ensemble was measured, from which a temperature was extracted using an analytical approach commonly used in the ultracold atom community.  Numerical trajectory simulations were then used to show the validity of the analytical approach.  The simulations offered deeper insight into the dynamics of the molecular ensemble and were subsequently used to investigate the effect of trap depth on the temperature of the molecules trapped on the molecule chip. This analysis allowed us to then use a phase-space manipulation process to significantly reduce the temperature of the trapped molecules, in this case to a third of its initial value.

\begin{acknowledgments}
We gratefully acknowledge the work of the electronic laboratory of the
Fritz Haber Institute, in particular Georg 
Heyne, Viktor Platschkowski and Thomas Zehentbauer. This work has been funded by the
European Community's Seventh Framework Program FP7/2007-2013 under
grant agreement 216 774 and ERC-2009-AdG under grant agreement
247142-MolChip. 
\end{acknowledgments}


\providecommand{\bysame}{\leavevmode\hbox to3em{\hrulefill}\thinspace}

\end{document}